\documentclass{appolb}

\usepackage{amsmath,amssymb,amsopn}
\usepackage{graphicx}

\newcommand{\be}{\begin{equation}}
\newcommand{\ee}{\end{equation}}
\newcommand{\ba}{\begin{eqnarray}}
\newcommand{\ea}{\end{eqnarray}}
\newcommand{\ban}{\begin{eqnarray*}}
\newcommand{\ean}{\end{eqnarray*}}
\newcommand \nn {\nonumber}

\begin{document}
\title{Parton Energy Loss \\ in an Unstable Quark-Gluon Plasma\thanks{Presented by 
K. Deja at the HIC-for-FAIR Workshop \& XXVIII Max-Born Symposium `Three Days on 
Quarkyonic Island', Wroc\l aw, Poland, May 18-21, 2011.}
}
\author{Margaret E. Carrington
\address{Department of Physics, Brandon University,\\
Brandon, Manitoba, R7A 6A9 Canada}
\and
Katarzyna Deja
\address{National Center for Nuclear Research, 00-681 Warsaw, Poland}
\and
Stanis\l aw Mr\' owczy\' nski
\address{Institute of Physics, Jan Kochanowski University, \\
25-406 Kielce, Poland \\
and National Center for Nuclear Research, 00-681 Warsaw, Poland}
}
\date{November 18, 2011}
\maketitle

\begin{abstract}

The energy loss of a fast parton scattering elastically in a weakly coupled quark-gluon 
plasma is formulated as an initial value problem. The approach is designed to study an 
unstable plasma, but it reproduces the well known result in the case of an equilibrium 
plasma. Contributions to the energy loss due to unstable modes are shown to exponentially 
grow in time. An unstable two-stream system is considered as an example. 

\end{abstract}

\section{Introduction}

When a highly energetic parton travels through the quark-gluon plasma (QGP), it losses its 
energy due to elastic interactions with plasma constituents. This is the so-called 
{\em collisional energy loss} which for the equilibrium QGP is well understood, see the 
review \cite{Peigne:2008wu} and the handbook \cite{lebellac}. The quark-gluon plasma produced 
in relativistic heavy-ion collisions, however, reaches the state of local equilibrium only after 
a short but finite time interval, and during this period the momentum distribution of plasma 
partons is anisotropic. Collisional energy loss has been computed for such an anisotropic 
QGP \cite{Romatschke:2004au}. However, the plasma with anisotropic momentum distribution 
is unstable due to the chromomagnetic modes (for a review see \cite{Mrowczynski:2005ki}), 
and the fact that unstable systems are explicitly time dependent as unstable modes exponentially
grow in time has not been taken into account in the study \cite{Romatschke:2004au}.

Our aim is to formulate an approach where the energy loss is found as the solution of an initial 
value problem. The parton is treated as an energetic classical particle with ${\rm SU}(3)$ color
charge. For the equilibrium plasma we recover the known results but for the unstable plasma 
the energy loss is shown to strongly depend on time.  Our approach to the energy-loss problem 
is similar to the method used earlier to study the momentum broadening $\hat{q}$ of a fast parton 
in anisotropic plasma \cite{Majumder:2009cf}. Analogous methods have also been used to study 
the spectra of chromodynamic fluctuations of an unstable plasma \cite{Mrowczynski:2008ae} which, 
in particular,  are responsible for the collisions integrals of transport equations \cite{Mrowczynski:2009gf}. 

Throughout the paper we use the natural system of units with $c=\hbar =k_B=1$ and the
signature of our metric tensor is $(+,-,-,-)$.

\section{General formula}
\label{general}

We consider a classical parton which moves across a quark-gluon plasma.  Its motion is 
described by the Wong equations \cite{Wong:1970fu}
\ba
\label{EOM-1a}
\frac{d x^\mu(\tau)}{d \tau} &=& u^\mu(\tau ) ,
\\
\label{EOM-1b}
\frac{d p^\mu(\tau)}{d \tau} &=& g Q^a(\tau ) \, F_a^{\mu \nu}\big(x(\tau )\big)
\, u_\nu(\tau ) ,
\\
\label{EOM-1c}
\frac{d Q_a(\tau)}{d \tau} &=& - g f^{abc} u_\mu (\tau ) \,
A^\mu _b \big(x(\tau )\big) \,
Q_c(\tau) ,
\ea
where $\tau$, $x^\mu(\tau )$, $u^\mu(\tau)$ and  $p^\mu(\tau)$ are, respectively, the parton's  
proper time, its trajectory, four-velocity and  four-momentum; $F_a^{\mu \nu}$ and $A_a^\mu$ 
denote the chromodynamic field strength tensor and four-potential along the parton's trajectory
and $Q^a$ is the classical color charge of the parton; $g$ is the coupling constant and
$\alpha_s \equiv g^2/4\pi$ is assumed to be small.  We also assume that the potential 
vanishes along the parton's trajectory  {\it i.e.} our gauge condition is 
$u_\mu (\tau ) \, A^\mu _a \big(x(\tau )\big) = 0 $. Consequently, due to Eq.~(\ref{EOM-1c})
the classical parton's  charge $Q_c(\tau)$ is constant within the chosen gauge.

The energy loss is given directly by Eq. (\ref{EOM-1b}) with $\mu = 0$. Using the 
time $t=\gamma\tau$ instead of the proper time $\tau$ and replacing the strength tensor 
$F_a^{\mu \nu}$ by the chromoelectric ${\bf E}_a(t,{\bf r})$ and chromomagnetic 
${\bf B}_a(t,{\bf r})$ fields, Eq.~(\ref{EOM-1b}) gives
\be
\label{e-loss-1}
\frac{dE(t)}{dt} = g Q^a {\bf E}_a(t,{\bf r}(t)) \cdot {\bf v} ,
\ee
where ${\bf v}$ is the parton's velocity. Since we consider a parton which is very energetic, 
${\bf v}$ is assumed to be constant and ${\bf v}^2 =1$.  Introducing the current generated by 
the parton ${\bf j}_a(t,{\bf r}) = g Q^a {\bf v} \delta^{(3)}({\bf r} - {\bf v}t)$, Eq.~(\ref{e-loss-1}) 
gives
\be
\label{e-loss-2}
\frac{dE(t)}{dt} = \int d^3r {\bf E}_a(t,{\bf r}) \cdot {\bf j}_a(t,{\bf r})  .
\ee

Since we deal with an initial value problem, we apply to the field and current not the usual 
Fourier transformation but  the so-called {\it one-sided Fourier transformation} defined as
\be
\label{1side}
f(\omega,{\bf k}) = \int_0^\infty dt \int d^3r
e^{i(\omega t - {\bf k}\cdot {\bf r})}
f(t,{\bf r}) ,
\ee
with the inverse 
\be
\label{1side-back}
f(t,{\bf r}) = \int_{-\infty +i\sigma}^{\infty +i\sigma}
{d\omega \over 2\pi} \int {d^3k \over (2\pi)^3}
e^{-i(\omega t - {\bf k}\cdot {\bf r})} f(\omega,{\bf k}) ,
\ee
where the real parameter $\sigma > 0$ is chosen is such a way that the integral over $\omega$ 
is taken along a straight line in the complex $\omega-$plane, parallel to the real axis, above all 
singularities of $f(\omega,{\bf k})$. Using Eqs. (\ref{1side}) and (\ref{1side-back}), 
Eq.~(\ref{e-loss-2}) can be rewritten:
\be
\label{e-loss-3}
\frac{dE(t)}{dt} = g Q^a
\int_{-\infty +i\sigma}^{\infty +i\sigma}
{d\omega \over 2\pi}
\int {d^3k \over (2\pi)^3}
e^{-i(\omega - \bar\omega)t} \; {\bf E}_a(\omega,{\bf k}) \cdot {\bf v} ,
\ee
where $\bar\omega \equiv {\bf k} \cdot {\bf v}$. 

The next step is to compute the chromoelectric field ${\bf E}_a$. Applying the one-sided 
Fourier transformation to the linearized Yang-Mills equations we get the set of equations 
familiar from electrodynamics 
\ba
\nonumber
i k^i \varepsilon^{ij}(\omega, {\bf k})
E^j_a(\omega, {\bf k})
&=& \rho_a (\omega,{\bf k}),
\;\;\;\;\;\;\;\;\;\;\;\;\;
i k^i B^i_a (\omega,{\bf k}) = 0 \;,
\\[2mm] 
\label{Maxwell-eqs-k}
i \epsilon^{ijk} k^j E_a^k(\omega,{\bf k})
&=& i\omega B_a^i (\omega,{\bf k}) + B_{0a}^i({\bf k}) ,
\\[2mm] \nonumber
i \epsilon^{ijk} k^j B_a^k (\omega,{\bf k})
&=& j_a^i(\omega,{\bf k})
-i\omega \varepsilon^{ij} (\omega,{\bf k})
E_a^j(\omega,{\bf k}) - D_{0a}^i({\bf k}) ,
\ea
where $ \rho_a$ is the color-charge density, the fields with the index 0 are 
initial values; the chromoelectric induction ${\bf D}_a$ is expressed as 
$D^i_a(\omega, {\bf k}) = \varepsilon^{ij}(\omega, {\bf k}) E^j_a(\omega, {\bf k})$
with $\varepsilon^{ij}(\omega, {\bf k})$ being chromodielectric tensor which carries 
all information about the medium. For an anisotropic plasma it equals
\begin{equation*}
\varepsilon^{ij} (\omega,{\bf k})
=  \delta^{ij} + 
{g^2 \over 2 \omega} \int {d^3 p \over (2\pi )^3}
{ v^i \over \omega - {\bf k} \cdot {\bf v} + i0^+} 
{\partial f({\bf p}) \over \partial p^k} 
\Big[ \Big( 1 - {{\bf k} \cdot {\bf v} \over \omega} \Big) \delta^{kj}
+ {k^k v^j \over \omega} \Big] ,
\end{equation*}
where $ f({\bf p})$ is the momentum distribution of plasma constituents. The color 
indices $a,b$ are dropped as $\varepsilon (\omega, {\bf k})$ is a unit matrix 
in color space. 

Although equations (\ref{Maxwell-eqs-k}) strongly resemble those of electrodynamics,
the gluon contribution to the color charge density $\rho_a$ and color current ${\bf j}_a$, 
which is a genuine nonAbelian effect, is fully incorporated in these equations.

Using Eq. (\ref{Maxwell-eqs-k}), the field $E^i_a(\omega, {\bf k})$ 
is found to be
\be
\label{E-field-k}
E^i_a(\omega, {\bf k}) = -i
(\Sigma^{-1})^{ij}(\omega,{\bf k})
\Big[ \omega j_a^j(\omega,{\bf k})
+ \epsilon^{jkl} k^k B_{0a}^l ({\bf k})
- \omega D_{0a}^j ({\bf k}) \Big] ,
\ee
where 
\be
\label{matrix-sigma}
\Sigma^{ij}(\omega,{\bf k}) \equiv
- {\bf k}^2 \delta^{ij} + k^ik^j
+ \omega^2 \varepsilon^{ij}(\omega,{\bf k}) .
\ee
Substituting the expression (\ref{E-field-k}) into Eq.~(\ref{e-loss-3}), we get the final formula
\ba
\label{e-loss-6}
\frac{dE(t)}{dt} &=& g Q^a v^i \int_{-\infty +i\sigma}^{\infty +i\sigma}
{d\omega \over 2\pi i}
\int {d^3k \over (2\pi)^3}
e^{-i(\omega -\bar{\omega})t}
(\Sigma^{-1})^{ij}(\omega,{\bf k})
\\ [2mm]\nn
&\times&
\Big[ 
\frac{i \omega g Q^a v^j}{\omega - \bar{\omega}}
+ \epsilon^{jkl} k^k B_{0a}^l ({\bf k})
- \omega D_{0a}^j ({\bf k}) \Big] .
\ea
The integral over $\omega$ is controlled by the poles of the matrix $\Sigma^{-1}(\omega,{\bf k})$  
which determine the collective modes in the system. Equivalently, these modes are found as 
solutions of the equation ${\rm det}[\Sigma(\omega,{\bf k})] =0$. Equation (\ref{e-loss-6}) needs 
to be treated differently for stable and for unstable systems.

\section{Stable systems}

When the plasma is stable, all modes are damped and the poles of $\Sigma^{-1}(\omega,{\bf k})$ 
are located in the lower half-plane of complex $\omega$. Consequently, the contributions to the 
energy loss corresponding to the poles of $\Sigma^{-1}(\omega,{\bf k})$ exponentially decay in time. 
The only stationary contribution is  given by the pole $\omega = \bar{\omega} \equiv {\bf k}\cdot {\bf v}$
which comes from the current ${\bf j}_a(\omega,{\bf k})$. Therefore, the terms in Eq.~(\ref{e-loss-6}), 
which depend on the initial values of the fields, can be neglected, and the energy loss of a fast parton 
in a stable plasma is 
\ba
\label{e-loss-stable-ave}
\frac{d \overline{E}(t)}{dt} &=& -i g^2 C_R v^i v^j
\int {d^3k \over (2\pi)^3} \; \bar{\omega} \;
 (\Sigma^{-1})^{ij}(\bar{\omega},{\bf k}) ,
\ea
where the bar indicates that averaging over parton's color state has been performed, and the factor 
$C_R$ is 4/3 for a quark and 3 for a gluon.

When the plasma is isotropic, the dielectric tensor can be expressed in a standard way through its 
longitudinal $\big(\varepsilon_L(\omega,{\bf k})\big)$ and transverse 
$\big(\varepsilon_T(\omega,{\bf k})\big)$ components and the matrix $\Sigma^{ij}(\omega,{\bf k})$
(\ref{matrix-sigma}) can be easily inverted. The energy loss (\ref{e-loss-stable-ave}) then equals
\ba
\label{e-loss-isotropic}
\frac{d \overline{E}}{dt} = -i g^2 C_R 
\int {d^3k \over (2\pi)^3} \;
\frac{\bar{\omega}}{{\bf k}^2} \;
\bigg[
\frac{1}
{\varepsilon_L(\bar{\omega},{\bf k})}
+ \frac{{\bf k}^2{\bf v}^2 - \bar{\omega}^2}
{\bar{\omega}^2
\varepsilon_T(\bar{\omega},{\bf k})-{\bf k}^2}
\bigg] ,
\ea
which corresponds to the standard energy loss due to soft collisions \cite{lebellac}.

\section{Unstable systems}

When the plasma is unstable, the matrix $\Sigma^{-1}(\omega,{\bf k})$ contains poles 
in the upper half-plane of complex $\omega$, and the contributions to the energy loss from 
these poles grow exponentially in time. After a sufficiently long time the parton's energy loss 
will be dominated by the fastest unstable mode. For an unstable plasma, the terms in 
Eq.~(\ref{e-loss-6}),  which depend on the initial values of fields ${\bf D}$ and ${\bf B}$, cannot 
be neglected, as these terms are amplified by a factor exponentially growing in time. Using 
Eq. (\ref{Maxwell-eqs-k}) the initial values can be computed as
\be
\label{D_0-2}
D^i_{0a}({\bf k}) = -i g Q^a \bar{\omega} \,
 \varepsilon^{ij}(\bar{\omega},{\bf k})
(\Sigma^{-1})^{jk}(\bar{\omega},{\bf k})  v^k ,
\ee
\be
\label{B_0}
B^i_{0a}({\bf k}) 
=\int_{-\infty}^{\infty} \frac{d\omega}{2 \pi} \,
 \frac{1}{\omega}\big({\bf k} \times {\bf E}_a(\omega, {\bf k})\big)^i
= -i g Q^a  \epsilon^{ijk}k^j
(\Sigma^{-1})^{kl}(\bar{\omega},{\bf k})  v^l.
\ee
Substituting Eqs. (\ref{D_0-2}) and (\ref{B_0}) into Eq.~(\ref{e-loss-6}) and averaging 
over the parton's color as before we obtain:
\ba
\label{e-loss-unstable}
&& \frac{d\overline{E}(t)}{dt} = g^2 C_R 
v^i v^l \int_{-\infty +i\sigma}^{\infty +i\sigma}
{d\omega \over 2\pi}
\int {d^3k \over (2\pi)^3}
e^{-i(\omega - \bar{\omega}) t}
(\Sigma^{-1})^{ij}(\omega,{\bf k})
\\ \nn
&&\times
\Big[ 
\frac{\omega \delta^{jl}}{\omega - \bar{\omega}}
-(k^j k^k - {\bf k}^2 \delta^{jk})
(\Sigma^{-1})^{kl}(\bar{\omega},{\bf k}) 
+ \omega \, \bar{\omega} \, \varepsilon^{jk}(\bar{\omega},{\bf k})
(\Sigma^{-1})^{kl}(\bar{\omega},{\bf k})   
 \Big] .
\ea
Equation~(\ref{e-loss-unstable}) gives the energy loss of a fast parton flying across the unstable 
plasma. The key point is that the presence of an unstable mode will produce a time dependent 
exponential growth of the form  $e^{{\rm Im} \omega({\bf k})t}$.

\section{Two-stream system}
\label{sec-2-streams}

In order to calculate the energy loss, one must invert the matrix $\Sigma^{ij}(\omega,{\bf k})$ 
defined by Eq.~(\ref{matrix-sigma}) to substitute the resulting expression into 
Eq. (\ref{e-loss-unstable}). For a general anisotropic system this is a tedious calculation, 
and therefore we consider for simplicity the example of the two-stream system which has 
unstable longitudinal electric modes. We assume that the chromodynamic field is dominated 
by the longitudinal chromoelectric field and take ${\bf B}(\omega, {\bf k}) = 0$ and  
${\bf E}(\omega, {\bf k}) = {\bf k} \big({\bf k}\cdot {\bf E}(\omega, {\bf k})\big) / {\bf k}^2$.
Then, $\Sigma^{ij}(\omega,{\bf k})$ is trivially inverted as
\be
\label{inv-sigma}
(\Sigma^{-1})^{ij}(\omega,{\bf k}) = 
\frac{1}{\omega^2 \varepsilon_L(\omega,{\bf k})}
\frac{k^ik^j}{{\bf k}^2}\,,~~
\varepsilon_L(\omega,{\bf k}) \equiv
\varepsilon^{ij}(\omega,{\bf k}) \frac{k^ik^j}{{\bf k}^2},
\ee
and Eq. (\ref{e-loss-unstable}) simplifies to 
\ba
\label{e-loss-2-stream-1}
\frac{d\overline{E(t)}}{dt} &=& 
 g^2 C_R 
\int_{-\infty +i\sigma}^{\infty +i\sigma}
{d\omega \over 2\pi}
\int {d^3k \over (2\pi)^3}
\frac{e^{-i(\omega - \bar{\omega}) t}}
{\omega^2 \varepsilon_L(\omega,{\bf k})}
\frac{\bar{\omega}^2}{{\bf k}^2}
\Big[ 
\frac{\omega}{\omega - \bar{\omega}}
+   \frac{\bar{\omega}}{\omega}
\Big] .
\ea
Eq.~(\ref{e-loss-2-stream-1}) gives a non-zero energy loss in the vacuum limit when 
$\varepsilon_L \rightarrow 1$.  Therefore, we subtract from the formula (\ref{e-loss-2-stream-1}) 
the vacuum contribution, or equivalently we replace $1/\varepsilon_L$ by  $1/\varepsilon_L - 1$.

The next step is to calculate $\varepsilon_L(\omega,{\bf k})$. With the distribution 
function of the two-stream system in the form
\be
\label{f-2-streams}
f({\bf p}) = (2\pi )^3 n 
\Big[\delta^{(3)}({\bf p} - {\bf q}) + \delta^{(3)}({\bf p} + {\bf q}) \Big] ,
\ee
where $n$ is the effective parton density in a single stream, one finds \cite{Mrowczynski:2008ae} 
\be
\label{eL-2-streams}
\varepsilon_L(\omega,{\bf k}) 
= \frac{\big(\omega - \omega_+({\bf k})\big)
\big(\omega + \omega_+({\bf k})\big)
\big(\omega - \omega_-({\bf k})\big)
\big(\omega + \omega_-({\bf k})\big)}
{\big(\omega^2 - ({\bf k} \cdot {\bf u})^2\big)^2} ,
\ee
where ${\bf u} \equiv {\bf q}/E_{\bf q}$ is the stream velocity, $\mu^2 \equiv g^2n/2 E_{\bf q}$ 
is a parameter analogous to the Debye mass squared, and $\pm \omega_{\pm}({\bf k})$ 
are the four roots of the dispersion equation 
$\varepsilon_L(\omega,{\bf k}) = 0$ which are
\ba
\label{roots}
\omega_{\pm}^2({\bf k}) &=& \frac{1}{{\bf k}^2}
\Big[{\bf k}^2 ({\bf k} \cdot {\bf u})^2
+ \mu^2 \big({\bf k}^2 - ({\bf k} \cdot {\bf u})^2\big)
\\[2mm] \nonumber
&\pm& \mu \sqrt{\big({\bf k}^2 - ({\bf k} \cdot {\bf u})^2\big)
\big(4{\bf k}^2 ({\bf k} \cdot {\bf u})^2 +
\mu^2 \big({\bf k}^2 - ({\bf k} \cdot {\bf u})^2\big)\big)} 
\; \Big] .
\ea
It is easy to see that $0 < \omega_+({\bf k}) \in \mathbb{R}$ for any ${\bf k}$.  
For ${\bf k}^2 ({\bf k} \cdot {\bf u})^2 \ge 2 \mu^2 \big({\bf k}^2 - ({\bf k} \cdot {\bf u})^2\big)$,
the minus mode is also stable, $0 < \omega_-({\bf k}) \in \mathbb{R}$, but for  
${\bf k} \cdot {\bf u} \not= 0$ and 
${\bf k}^2 ({\bf k} \cdot {\bf u})^2 < 2 \mu^2 \big({\bf k}^2 - ({\bf k} \cdot {\bf u})^2\big)$ one finds 
that $\omega_-({\bf k})$ is imaginary which is the well-known two-stream electric instability. 

Strictly speaking, the stream velocity ${\bf u}$ given by the distribution function (\ref{f-2-streams}) 
equals the speed of light. However, the distribution (\ref{f-2-streams}) should be treated
as an idealization of a two-bump distribution with bumps of finite width. Then, the momenta 
of all partons are not  exactly parallel or antiparallel and  the velocity ${\bf u}$, which enters 
Eqs.~(\ref{eL-2-streams}, \ref{roots}), obeys ${\bf u}^2 \le 1$. 

Equations (\ref{e-loss-2-stream-1}) and (\ref{eL-2-streams}) determine the energy loss of 
a parton in the two-stream system.  The integral over $\omega$ can be computed analytically 
as it determined by the six poles of the integrand located at  
$\omega = \pm \omega_+({\bf k}), \; \pm \omega_-({\bf k}), \; \bar\omega$ and $0$. The 
remaining integral over ${\bf k}$ must be done numerically. 

\begin{figure*}[t]
\centering
\includegraphics[width=10cm]{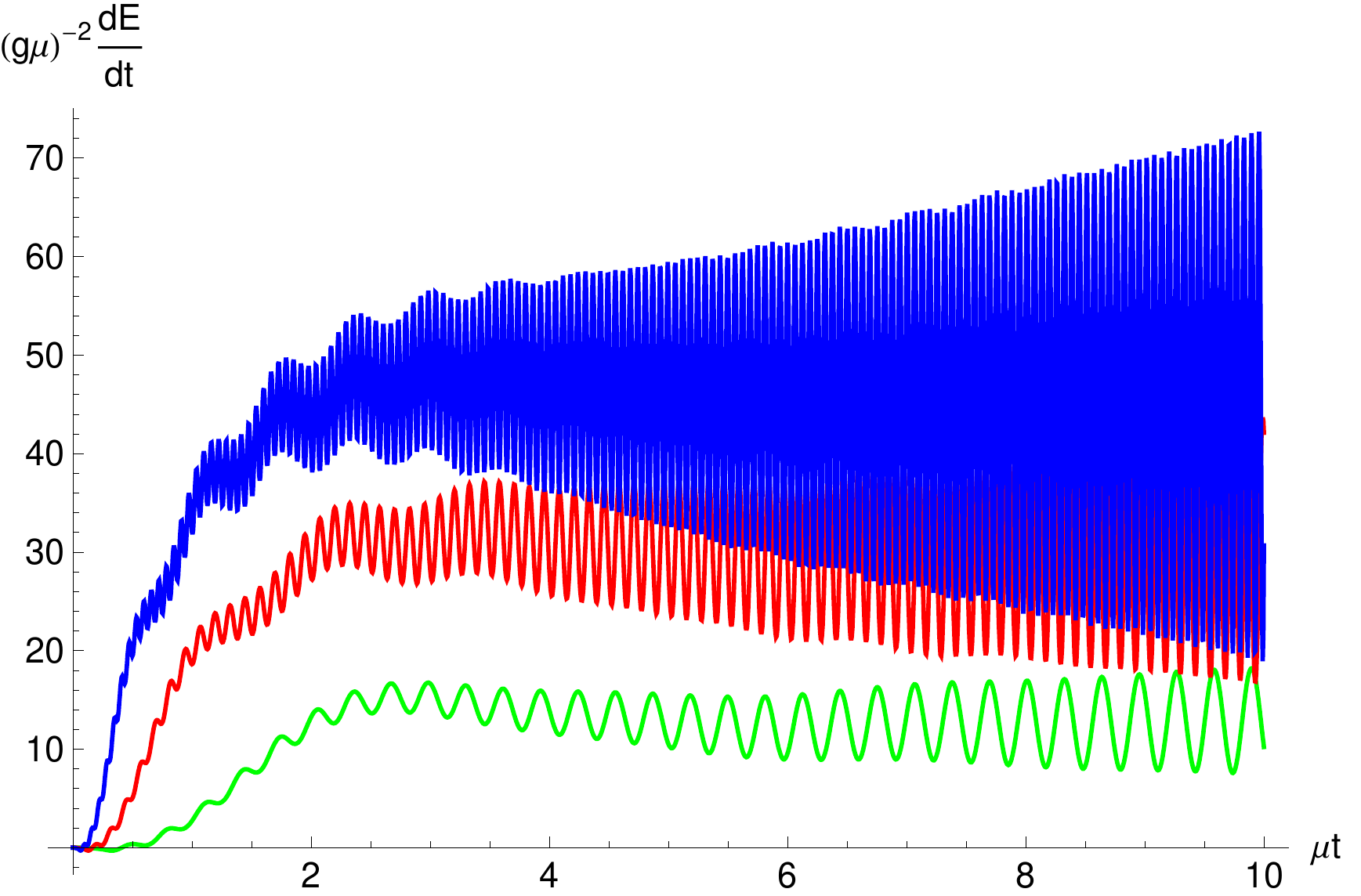}
\caption{The energy loss per unit length as a function of time for a parton flying  along 
the direction of streams. The lowest (green) curve corresponds to $k_{\rm max} = 20$, 
the middle (red) one to $k_{\rm max} = 50$ and the most upper (blue) curve to 
$k_{\rm max} = 100$.}
\label{fig-e-loss}
\end{figure*}

Performing the calculations we have redefined all dimensional quantities by multiplying 
them by the appropriate power of $\mu$ to obtain dimensionless variables. We have 
chosen the following values of the parameters: $g=1$, $|{\bf v}| = 1$, $|{\bf u}| = 0.9$, 
$C_R = 3$. To take the integral over ${\bf k}$, cylindrical coordinates with the axis 
$z$ along the stream velocity ${\bf u}$ have been used. Since the integral 
appears to be divergent,  it has been taken over a finite domain such that  
$-k_{\rm max} \le k_L \le k_{\rm max}$ and  $0 \le k_T \le k_{\rm max}$. In Fig.~\ref{fig-e-loss} 
we show the parton's energy loss per unit length as a function of time for a parton flying  
along the streams for $k_{\rm max} = 20,\; 50, \;100$. The energy loss strongly oscillates 
with amplitude growing in time. However, the results depend on the cut-off parameter  
$k_{\rm max}$: the magnitude of energy loss grows and the period of oscillations shrinks 
when $k_{\rm max}$ increases. This ultraviolet sensitivity  of our results is not very surprising, 
as our approach is fully classical. In the case of equilibrium (stable) plasma, the energy loss due 
to soft interactions diverges logarithmically with $k_{\rm max}$ \cite{Peigne:2008wu}. The 
divergence signals a necessity to combine the classical contribution to the energy loss at small 
wave vectors with the quantum contribution at higher ones. A quantum approach to the parton 
energy loss in unstable plasma needs to be developed. 

Although the two-stream system is not directly relevant to the quark-gluon plasma which is produced 
in relativistic heavy-ion collisions, let us comment on the potential interest of our study. Choosing 
$\mu =$ 200 MeV, which is a rough estimate of Debye mass in the QGP from nuclear collisions, 
the units on the horizontal and vertical axis in Fig.~\ref{fig-e-loss} are, respectively, fm and 200 MeV/fm. 
Since the parton's energy loss in the QGP observed in relativistic heavy-ion collisions is of order 1 GeV/fm, 
the magnitude of the energy loss discussed here is certainly of phenomenological interest. 

\section{Conclusions}

We have developed a formalism where the energy loss of a fast parton in a plasma medium
is found as the solution of initial value problem. The formalism allows one to obtain the energy
loss in unstable plasma where some modes exponentially grow in time. In the case of stable 
plasma, one reproduces correctly the standard energy-loss formula. As an example of an
unstable system we have studied a two-stream system. The energy loss per unit length
is not constant, as in an equilibrium plasma, but it exhibits strong time dependence.

\section*{Acknowledgments}

This work was partially supported by Polish Ministry of Science and Higher Education under 
grants N~N202~204638 and 667/N-CERN/2010/0.


\end{document}